\documentclass[twocolumn,amssymb,amsmath,showpacs,aps,pre,10pt,superscriptaddress]{revtex4-1}
\usepackage[dvips]{graphicx}
\usepackage{epsfig}

\begin{document}

\title{Narrow-escape-time problem: the imperfect trapping case}

\author{F{\'e}lix Rojo}

\affiliation{Fa.M.A.F., Universidad Nacional de C\'ordoba, Ciudad Universitaria, X5000HUA C\'ordoba, Argentina}

\author{Horacio S. Wio}

\affiliation{Instituto de F\'{\i}sica de Cantabria, Universidad de
Cantabria and CSIC, E-39005 Santander, Spain}

\author{Carlos E. Budde}
\affiliation{Fa.M.A.F., Universidad Nacional de C\'ordoba, Ciudad Universitaria, X5000HUA C\'ordoba, Argentina}

\begin{abstract}
We present a master equation approach to the \emph{narrow escape time}
(NET) problem, i.e. the time needed for a particle contained in
a confining domain with a single narrow opening, to exit
the domain for the first time. We introduce a finite transition
probability, $\nu$, at the narrow escape window allowing the study
of the imperfect trapping case. Ranging from $0$ to $\infty$, $\nu$ allowed the study of both extremes of the trapping process: that of a highly deficient capture, and situations where escape is certain (``perfect trapping'' case).
We have obtained analytic results for the basic quantity
studied in the NET problem, the \emph{mean escape time} (MET), and we have
studied its dependence in terms of the transition (desorption) probability
over (from) the surface boundary, the confining domain dimensions, and the
finite transition probability at the escape window. Particularly we show
that the existence of a global minimum in the NET depends on the
`imperfection' of the trapping process. In addition to our analytical
approach, we have implemented Monte Carlo simulations, finding excellent
agreement between the theoretical results and simulations.
\end{abstract}


\maketitle

\section{Introduction}

The time needed for a particle contained in a confining domain with a single
small opening to exit the domain for the first time, usually
referred as \emph{narrow escape time} problem (NET), finds a
prominent place in many domains and fields. For instance in cellular biology,
it is related to the random time needed by a particle
released inside a cell to activate a given mechanism on the cell membrane (\cite{020,030,040}).
Generally speaking the NET problem is part of the so called \emph{Intermittent processes}, which are used to explain scenarios
ranging from animal search patterns (\cite{000}), through the solutions
or melts of synthetic macromolecules (\cite{055,060}), to the manufacture of self-assembled mono- and multi-layers (\cite{065,066}).
Since the work of Berg and Purcell (\cite{070}), research in the
NET problem area has experienced a steady growth over time and motivated a great deal of work (\cite{071,073,075,078,079,*080,*090,095,100,110,114,*115,210,211,212}).

In \cite{210}, we have introduced an analytical Markovian model that showed
the impact of geometrical parameters and the interplay between surface and
boundary paths in the studied confining domain, of a discrete and rectangular
shaped nature, for the perfect trapping case. With ``perfect trapping'' we refer
to the particle's impossibility of return to the system, i.e. once the particle
reaches the narrow opening the escape becomes certain. In that work we presented
a phase diagram which showed that some combinations of the geometrical parameter
and the transport mechanism were required for the existence of an optimal transport
(a global minimum in the NET).

In this work we consider the same confining domain and the transport properties
that we dealt with in Ref. \cite{210}. However we eliminate the assumption of
perfect escape introducing a finite transition probability at the narrow escape
window. It is well known that systems description through the ``imperfect
trapping case'' (once in the trap site capture is not certain) are suitable
whenever the surface contains `deep traps', capture and re-emission from a
surface that contains sites with several internal states such as the `ladder
trapping model', proteins with active sites deep inside its matrix, etc.
(\cite{180,190,200,201,*202}). Under this new assumption, we have discovered some very
interesting results. Particularly we show that the existence of a global
minimum in the NET depends on the existence of an imperfection in the trapping
process.

The aim of this work is to study the influence of the `imperfection' in the
passage through the escape window on the effective diffusion process, and its
effect on the NET problem. For that purpose we calculate the time required by
the particle to escape from the system (MET). In  order to perform our research
we exploit Dyson's theory (\cite{050}), and the notions of \emph{Absorption
Probability Density} (APD) (\cite{190}).

The outline of this paper is as follows. In the following section we
introduce some general results regarding imperfect trapping process
as well as our model, and provide the basic definitions and concepts. We also
describe the proposed analytical approach and present the main results.
Section III depicts several assorted illustrations for the MET to the narrow
opening for different configurations of the system through a comparison between
our analytical framework and Monte Carlo simulations. In Section IV we discuss
our conclusions and perspectives. Finally, in the Appendix we present the
calculations indicated in Section II.

\section{Analytical Approach}

\subsection{Some general results regarding imperfect trapping}
\label{GeneralResults}

Let us consider the problem of a walker making a random walk in some
finite domain with a trap or sink present in the system. We will follow
the walker evolution through the system considering the `unrestricted'
conditional probability $P(\vec{s}, t|\vec{s}_0, t = 0)$, that is, the
probability that a walker is at $\vec{s}$ at time $t$ given it was at
$\vec{s}_0$ at $t = 0$. By `unrestricted' we identify a situation with
no traps/sinks present in the system.
\begin{figure}[htb]
\begin{center}
\includegraphics[clip,scale=0.9]{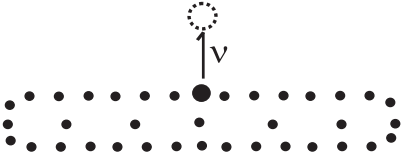}\vspace{0.0cm}
\caption{Finite domain with a trap or sink present. The entrance to the
trap site (empty circle) is regulated at the escape `window' by the
transition rate $\nu$.}
\label{Fig1}
\end{center}
\end{figure}
\subsubsection*{Absorption Probability Density and Mean Absorption Time}

As we are interested in the \emph{trapping process}, let us define $A(\vec{s},t|\vec{s}_0,0)$
as the \textit{Absorption (trapping) Probability Density} (APD) through the site $\vec{s}$ at
time $t$, given that the walker was at
$\vec{s}_0$ at time $t=0$, i.e., $A(\vec{s},t|\vec{s}_0,0)\,dt$ gives
the trapping probability of the walker, through $\vec{s}$, between
$t$ and $t + dt$ given that it started at $t=0$ from $\vec{s}_0$. It
is worth mentioning that the \textit{First Passage time} approach is not
fully applicable since an `excursion' to the trapping site does not
necessarily ends the process. However we show in the following lines
that an interesting relation could be established between $A(\vec{s},t
|\vec{s}_0,0)$ and $F(\vec{s},t|\vec{s}_0,0)$, the First Passage time
density (FPTD) through the
site $\vec{s}$ at time $t$, given that the walker was at $\vec{s}_0$
at time $t=0$.

From now on we will denote the \emph{Laplace} transform of a function $f(t)$ by its argument,
\begin{equation*}
\mathcal{L}\{f(t)\}=f(u)=\int_0^{\infty}e^{-ut}f(t) \,dt\,.
\end{equation*}

The connection between the APD and the `unrestricted' conditional
probability $P(\vec{s},t|\vec{s}_0,t=0)$ could be traced to results
in \cite{190} or \cite{220}. This approach in the \emph{Laplace}
domain gives,
\begin{equation}\label{APD1}
A(\vec{0},u|\vec{s}_0,t=0)=\frac{\nu P(\vec{0},u|\vec{s}_0,t=0)}
{1+\nu P(\vec{0},u|\vec{0},t=0)} \,.
\end{equation}

Let us make a brief digression regarding the relation between the APD and
the FPTD. For this we rewrite Eq. (\ref{APD1}) in the form
\begin{equation}\label{APD2}
A(\vec{0},u|\vec{s}_0,t=0)\!\!=\!\!\frac{P(\vec{0},u|\vec{s}_0,t=0)}{P(\vec{0},u|\vec{0},t=0)}\!\cdot\!\frac{\nu}{\nu+\frac{1}{P(\vec{0},u|\vec{0},t=0)}}
\end{equation}
As usual the connection between FPTD and the `unrestricted' conditional
probability $P(\vec{s},t|\vec{s}_0,t=0)$ is established (in the \emph{Laplace} domain)
through the `Renewal approach' (\cite{130}),
\begin{equation}\label{F1P}
F(\vec{0},u|\vec{s}_0,t=0)=
\frac{P(\vec{0},u|\vec{s}_0,t=0)}
{P(\vec{0},u|\vec{0},t=0)} \,.
\end{equation}
We recognize in equation (\ref{APD2}), the FPTD (\ref{F1P}) and using the
relation $F(\vec{0},u|\vec{0},t=0)=1-\Psi(\vec{0},u)P(\vec{0},u|\vec{0},t=0)^{-1}$,
where $\Psi(\vec{s},\tau)$ is the probability that the walker remains at $\vec{s}$
until time $\tau$ since it arrived at $\vec{s}$ at time $0$ (in the unrestricted case) \footnote{See \cite{180} and references therein.}, we rewrite Eq. (\ref{APD2}) as
\begin{eqnarray}\label{APD3}
\nonumber A(\vec{0},u|\vec{s}_0,t=0)&=&F(\vec{0},u|\vec{s}_0,t=0)\frac{\nu}{\Psi(\vec{0},u)^{-1}+\nu}\cdot\\
&& \cdot\frac{1}{1-\frac{\Psi(\vec{0},u)^{-1}}{\Psi(\vec{0},u)^{-1}+\nu}F(\vec{0},u|\vec{0},t=0)}
\end{eqnarray}
The term $\nu(\Psi(\vec{0},u)^{-1}+\nu)^{-1}$ in (\ref{APD3}) gives the fraction of
walkers that are trapped while $\Psi(\vec{0},u)(\Psi(\vec{0},u)^{-1}+\nu)^{-1}$
gives the ones that are not absorbed. Further considerations could be made
regarding (\ref{APD3}), we write it as,
\begin{equation}\label{sum_APD}
A(\vec{0},u|\vec{s}_0,t=0)=\sum_{j=1}^{\infty}A_{j}(\vec{0},u|\vec{s}_0,t=0)\, ,
\end{equation}
where
\begin{eqnarray}\label{APDj}
\nonumber A_{j}(\vec{0},u|\vec{s}_0,0)\!&=&\!F(\vec{0},u|\vec{s}_0,0)\!\!\left(\frac{F(\vec{0},u|\vec{0},0)\Psi(\vec{0},u)^{-1}}{\Psi(\vec{0},u)^{-1}+\nu}\right)^{j-1}\!\!\!\!\!\!\cdot\\ &&\cdot\left(\frac{\nu}{\Psi(\vec{0},u)^{-1}+\nu}\right)\,.
\end{eqnarray}
Notice that $A_{j}(\cdot)$ accounts for that walkers that are absorbed in the
$j$-visit ($j=2,3,\ldots$) and \emph{not before}, i.e., the walkers arrive for the
first time at site $\vec{0}$ from $\vec{s}_0$ but these are not absorbed until
they return to site $\vec{0}$ for the $(j-1)$-time.

The probability of being absorbed in the $j-$visit at site $\vec{0}$ is,
\begin{eqnarray}\label{APDj=0}
\int_0^{\infty}A_{j}(\vec{0},t|\vec{s}_0,0)dt\!&=& A_{j}(\vec{0},u=0|\vec{s}_0,0)\\ \nonumber&=&\!\!\!\left(\frac{\Psi(\vec{0},0)^{-1}}{\Psi(\vec{0},0)^{-1}+\nu}\right)^{j-1}\!\!\!\!\!\!\cdot\left(\frac{\nu}{\Psi(\vec{0},0)^{-1}+\nu}\right)\,,
\end{eqnarray}
where $\Psi(\vec{0},u=0)$ is the mean residence time at site $\vec{0}$ in the unrestricted
system and we have used that $F(\vec{s},u=0|\vec{s}_0,t=0)=1$ for a finite (unrestricted)
system. As equation (\ref{APDj=0}) shows, when $\nu\rightarrow0$ independently of the
$j$ value, we have no absorption, while in the limit $\nu\rightarrow\infty$ the absorption
is certain in the first `visit' to the site (i.e. $j=1$).

The mean absorption time, i.e., the mean time until the walker is absorbed is
evaluated in terms of $A(\cdot)$ as,
\begin{eqnarray}
\nonumber T&=&\int_0^{\infty}t\,\sum_{\vec{s}_0}A(\vec{0},t|\vec{s}_0,0)g(\vec{s}_0)\,dt\\
\label{Tdef}&=&-\frac{\partial}{\partial u}\left\{\sum_{\vec{s}_0}A(\vec{0},u|\vec{s}_0,0)g(\vec{s}_0)
\right\}\bigg|_{u=0}
\end{eqnarray}
where $g(\vec{s}_0)$ denotes the probability density of initially finding the walker
at a position $\vec{s}_0$ (\cite{130}).

\subsection{The Model}\label{TheModel}

\begin{figure}[htb]
\begin{center}
\includegraphics[clip,width=0.48\textwidth]{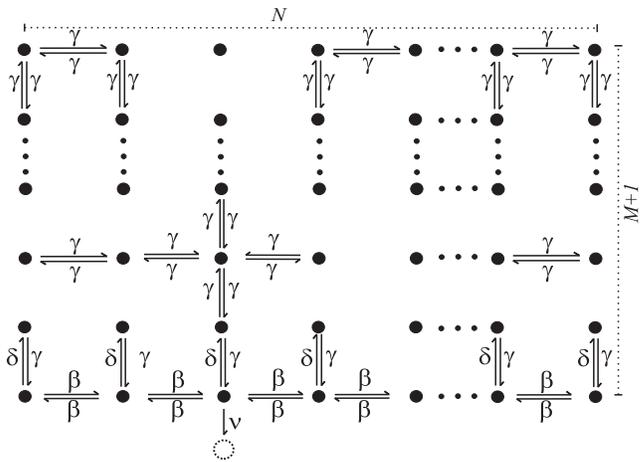}\vspace{0.0cm}
\caption{Schematic transitions of the walker to/from the base line and
to/from a generic surface site. Notice that the entrance to the trap/escape
site (empty circle) is regulated by the transition rate $\nu$ also notice
that it could be reached both from the surface (with transition rate $\gamma$)
and from the baseline (with transition rate $\beta$).}
\label{Fig2}
\end{center}
\end{figure}

For our model we consider the problem of a walker making a random walk
in a finite rectangular $N\times (M+1)$ lattice (see figure \ref{Fig2}).
The surface is bounded in the $y$ direction where the walkers can move
from $y = 0$ to $y = M$, and periodic boundary conditions are assumed
in the $x$ direction so $x$ and $x + N$ denote the same place in space.
As we mentioned before, we follow the walker's evolution through the
system considering the conditional probability
$P(n,m,t|n_0,m_0,t=0)\equiv P(n,m,t)$, where $(n, m)$ are discrete
coordinates in the $(x,y)$ space. $P(n,m,t)$ satisfies the following
master equation:
\begin{eqnarray}
\nonumber\dot{P}(n,0,t)&=&\gamma P(n,1,t)-\delta P(n,0,t)\\
\nonumber& &+\beta\big(P(n+1,0,t)+P(n-1,0,0,t)\\
\nonumber& &-2P(n,0,t)\big) \,; \qquad\qquad m=0\\
\nonumber\dot{P}(n,1,t)&=&\delta P(n,0,t)-4\gamma P(n,1,t)\\
\nonumber& &+\gamma\big(P(n+1,1,t)+P(n-1,1,t)\\
\nonumber& & + P(n,2,t)\big) \,; \qquad\qquad m=1\\
\nonumber\dot{P}(n,m,t)&=&\gamma\big(P(n-1,m,t)+P(n+1,m,t)\\
\nonumber& &+P(n,m+1,t)+P(n,m-1,t)\big)\\
\nonumber& &-4\gamma P(n,m,t) \, ; \qquad\qquad 2\leq m \leq M-1\\
\nonumber\dot{P}(n,M,t)&=&\gamma\big(P(n-1,M,t)+P(n+1,M,t)\\
\nonumber& &+ P(n,M-1,t)\big)\\
\nonumber& &-3\gamma P(n,M,t) \, ; \qquad\qquad m=M \\\label{MsEq1}
\end{eqnarray}
where $\gamma$ is the surface transition probability per unit time
in the $x$ and $y$ direction, $\beta$ is the transition probability
over the line $m=0$ in the $x$ direction, and $\delta$ is the
desorption probability per unit time from the boundary line $m=0$.

We introduce the imperfect escape case by allowing a finite transition
probability ($\nu$) at the narrow escape window. Varying from $0$ to
$\infty$, $\nu$ allowed the study of both a deficient trapping, and
situations where escape is certain (i.e. perfect trapping case).

In the following we will say that the walker `escapes' when
it gets trapped or adsorbed, without the possibility of returning
to the system. This terminology matches the one used
in the NET area. Hence, \emph{adsorption} $\rightarrow$ \emph{escape},
and so on.

\subsubsection*{Escape Probability Density (EPD)}

We now make a brief comment regarding the Escape Probability Density.
Taking into account the parameters of our model, we could write
 equation (\mbox{\ref{APDj=0}}) as,
\begin{eqnarray}\label{APDj=0_model}
\int_0^{\infty}A_{j}(\vec{0},t|\vec{s}_0,0)dt\!&=& A_{j}(\vec{0},u=0|\vec{s}_0,0)\\ \nonumber&=&\!\!\!\left(\frac{2\beta+\delta}{2\beta+\delta+\nu}\right)^{j-1}\!\!\!\!\!\!\cdot\left(\frac{\nu}{2\beta+\delta+\nu}\right)\,.
\end{eqnarray}
Notice that as $\nu$ grows ($\nu>>2\beta+\delta$) each $A_j$ becomes smaller
except for $A_1$, with $A_1\rightarrow1$, i.e. the escape is certain in the first visit. However when $\nu<<2\beta+\delta$, the probability of escape $A(\cdot)$  has contributions from each $j$-visit. This can best be understood considering,
\begin{equation}\label{Aj_coc}
\left.\frac{A_j(\cdot)}{A_{j+1}(\cdot)}\right|_{u=0}=1+\frac{\nu}{2\beta+\delta} \qquad (j\neq1)
\end{equation}
which gives the relative contribution of successive terms in (\ref{sum_APD}).
When $\nu$ gets smaller the contribution is spread all over $j$ values as
Eq. (\ref{Aj_coc}) shows. In contrast, each $A_j\rightarrow0$ ($j\neq1$) as
$\nu$ grows, concentrating all the probability in $A_1$.

\subsubsection*{Mean Escape Time (MET)}

Following the ideas exposed in \cite{210} and by resorting to the matrix
formalism and the Dyson's procedure (\cite{050}) we were able to obtain the
probability $P(n,m, t|n_0,m_0,t=0)$ (in the Fourier-Laplace space),
which is the building block for the MET. For the detailed
calculation see appendix \ref{AppendixA}.

We will denote the (finite) \emph{Fourier} transform by its argument
, as we did in the \emph{Laplace} transform case.
Thus for example the transform on a coordinate, say $x$, would read:
\begin{eqnarray}
P(k,m,t|n_0,m_0,0) &\equiv&
\mathcal{F}\left\{P(n,m,t|n_0,m_0,0)\right\}
\nonumber \\
&=& \sum_{n=0}^{N-1}e^{ikn}P(n,m,t|n_0,m_0,0)\,.
\nonumber
\end{eqnarray}
From $P(k,m, u|n_0,m_0,t=0)$ (obtained in the Fourier-Laplace space),
the probability that a walker is on the surface at site $(n, m)$ at
time $t$ given it was at $(n_0, m_0)$  at $t = 0$, $P(n,m,t|n_0,m_0,t=0)$,
is derived by using the inverse \emph{Laplace} transform on $u$ and the
inverse \emph{Fourier} transform on $k$ (for the $x$ coordinate) for
each $\left[\mathbb{P}(k,u)\right]_{m,m_0}$. However, as we are interested
in the calculation of (\ref{Tdef}), we only need to perform the inverse
Fourier transform on $P(0,0,u|n_0,m_0,t=0)$, i.e. we need the
elements $\mathcal{F}^{-1}\left\{\left[\mathbb{P}(k,u)\right]_{0,m_0}\right\}$.
We obtain for $\left[\mathbb{P}(k,u)\right]_{0,m_0}$:
\begin{eqnarray}
\nonumber\left[\mathbb{P}(k,u)\right]_{0,m_0}\!\!=\!\!\frac{\eta^{m_0}+\eta^{\tilde{M}-m_0}}
{\delta(1-\eta)(1-\eta^{\tilde{M}-1})+(u-A_1(k))(1+\eta^{\tilde{M}})}\, ,\\
\end{eqnarray}
where $\eta=1+\left(\tilde{u}-\sqrt{\tilde{u}^2+4\gamma\tilde{u}}\right)/2\gamma$,
$\tilde{M}=2M+1$ and $\tilde{u}=u-A(k)$. The inverse \emph{Fourier} transform on
$\left[\mathbb{P}(k,u) \right]_{0,m_0}$ is carried out in the following way,
\begin{equation}
P(0,0,u|n_0,m_0,0)\!=\!\frac{1}{N}\!\sum_{q=0}^{N-1}\!e^{i\frac{2\pi n_0
q}{N}}\big[\mathbb{P}(\frac{2\pi q}{N},u)\big]_{0,m_0}
\end{equation}

Thus we have obtained the required expression for the calculation of the MET
through the narrow escape window and it only remains to choose the initial distribution. We now evaluate the MET for a walker with an
uniform initial distribution on the base line ($y=0$). This means the initial distribution is given by $ g(n,m)=(1-\delta_{n,0})\delta_{m,0}/(N-1)$. Notice that we explicitly exclude the possibility of having a walker at $(0,0)$ at $t = 0$ \footnote{This consideration will help us in the comparison with the perfect escape case and avoids the `instantaneous' escaping in the limit
	$\nu\rightarrow\infty$}. This way we obtain,
\begin{equation}\label{Tsol}
T=\!\!N\left[\frac{M}{\gamma}+\frac{1}{\delta}\right]\!\!\left\{\frac{\delta}{\nu}+\frac{\delta}{N-1}\sum_{q=1}^{N-1}\left[\mathbb{P}(\frac{2\pi q}{N},u=0)\right]_{0,0}\right\}
\end{equation}
We make some comments regarding equation (\ref{Tsol}) which constitutes one of
our main results. Notice that (\ref{Tsol})  adequately provides the limits of
perfect escape case, $\nu\rightarrow\infty$ (obtained in \citep{210}) and no
escape window, $\nu\rightarrow0$,  T$\rightarrow\infty$. Observe that
as it is commented in \mbox{\cite{100}} for a perfect escape case, T could be expressed in the following way,
\begin{eqnarray}\label{Tsol_sep}
\nonumber T&=&\frac{NM}{\gamma}\left[\frac{\delta}{\nu}+\delta\frac{\sum_{q}\left[\mathbb{P}(\frac{2\pi q}{N},u=0)\right]_{0,0}}{N-1}\right]\\
&&+\frac{N}{2\beta}\left[\frac{2\beta}{\nu}+2\beta\frac{\sum_{q}\left[\mathbb{P}(\frac{2\pi q}{N},u=0)\right]_{0,0}}{N-1}\right]
\end{eqnarray}
or $T=T_{surface}+T_{line}$; \mbox{\eqref{Tsol_sep}} provides an interesting physical insight into the problem. Simply notice how the mean escape time is constructed from the mean duration of \emph{surface} excursions and the mean duration of \emph{border or line} excursions (first and second terms of (\ref{Tsol_sep}) respectively).

\subsubsection*{Regarding the existence of a minimum in T}

$T$ could be enhanced with respect to $\delta$ provided we are able to find $\delta^{*}=\delta^{*}(\beta,\gamma,\nu,N,M)$ - the optimal desorption
probability - that satisfies,
\begin{equation}\label{delta_min}
\left.\frac{\partial T}{\partial\delta}\right|_{\delta=\delta^*}\!\!\!\!=\frac{NM}{\gamma\nu}+ \frac{N}{N-1}\sum_{q=1}^{N}\frac{M\gamma^{-1}2\beta\alpha_2-\alpha_1}{(\alpha_1\delta^*+\alpha_{2}2\beta)^2}=0
\end{equation}
where the $\alpha_i=\alpha_i(q,N,M)$ for $i=1,2$ are defined in appendix
(\ref{AppendixB}). Notice that (\ref{delta_min}) defines an implicit equation
for $\delta^*$ which, although we could not solve, provided us with some generals conclusions as it approaches certain limits. Consider first $\beta\rightarrow0$, with the other parameters held fixed. In this case (a finite value for) $\delta^*$ exists whenever we have a finite escape probability rate $\nu$ and,
\begin{equation}\label{delta_min0}
\delta^*=\sqrt{\frac{\nu\gamma}{M(N-1)}\sum_{q=1}^{N-1}\alpha_1^{-1}}\,.
\end{equation}
This is a highly interesting result since in the perfect
escape case ($\nu=\infty$) this minimum disappears, as $\delta^*$ is pushed towards
$\infty$. Thus the `imperfect' escape window enables a region that was
absent in the perfect case. On the other hand in the
limit $\beta>>\delta$ it could be shown that equation (\ref{delta_min})
can not be satisfied for any $\delta^*$ value. In this
case, and taking into account the walker's initial distribution, the
transport is performed on the baseline (lower boundary) of the confining
domain, so this is an expected behavior.

In the following section we will make more remarks
regarding the minimum in T, while introducing
some illustrations corresponding to $\delta^*$.

\section{Illustrations}
\label{Illustrations}

Here we illustrate the general framework introduced in the previous
section and compare our theoretical results to independent Monte
Carlo simulations. In the next figures, lines indicate analytical
calculations while symbols correspond to Monte Carlo (MC) simulations.
We will be interested in situations in which a \emph{mixed} type of
transport generates a global minimum in the
\textit{Mean Escape Time}.
\begin{figure}[!htb]
\begin{center}
\includegraphics[clip,width=0.48\textwidth]{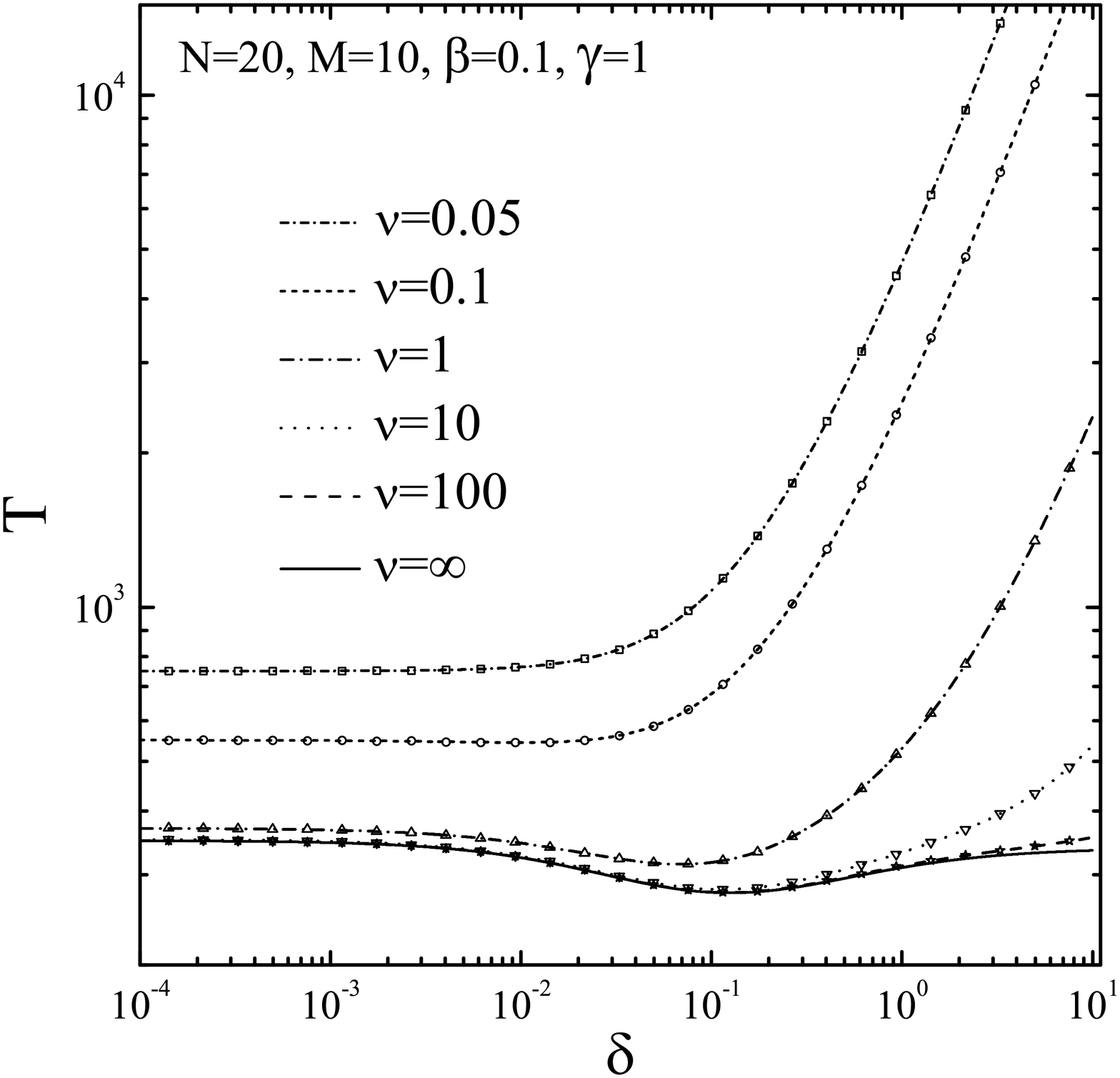}\vspace{0.0cm}
\caption{MET as a function of the desorption rate (in log scale) $\delta$,
with $M=10$, $N=20$, $\beta=0\textrm{.}1$, for different values of the
transition rate at the escape window $\nu$. From top to bottom $\nu=0.05,
0.1, 1, 10, 100$. We have also included (thick solid line) for comparison,
the \textit{perfect escape window} case ($\nu=\infty$).}
\label{Fig3}
\end{center}
\end{figure}

\begin{figure}[!htb]
\begin{center}
\includegraphics[clip,width=0.48\textwidth]{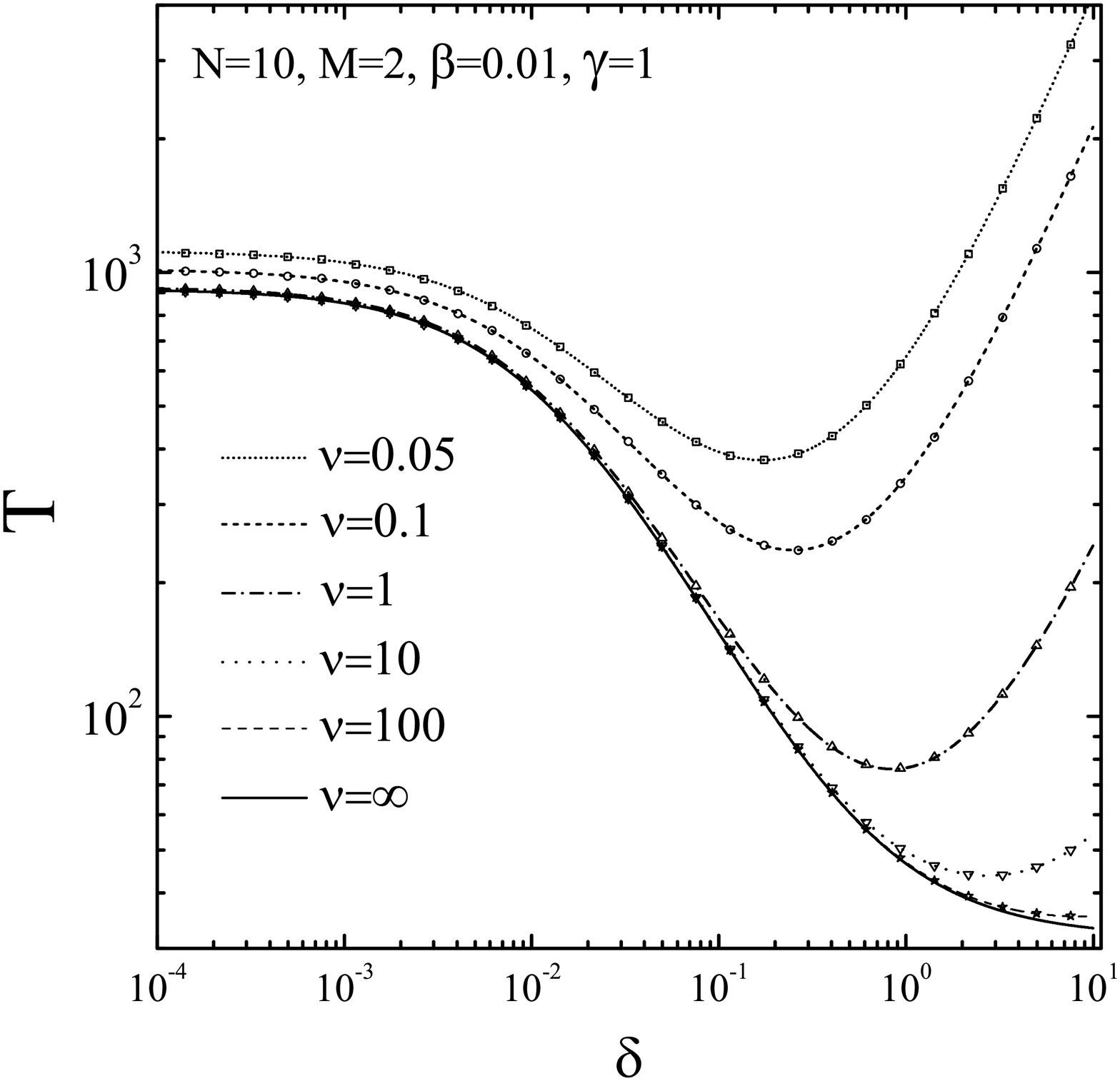}\vspace{0.0cm}
\caption{MET as a function of the desorption rate (in log-log scale)
$\delta$, with $\beta=0.01$, $N=10$, $M=2$, for different values of
$\nu$ (transition rate at the escape window). From bottom to top
$\nu=\infty, 100, 10, 1, 0\textrm{.}1, 0\textrm{.}05$. Lines correspond
to analytical calculations and symbols to Monte Carlo simulations.}
\label{Fig4}
\end{center}
\end{figure}

In Fig.~\ref{Fig3} we present curves corresponding to the MET (Mean Escape
Time), as a function of the desorption rate $\delta$, with parameters
$N=20$, $M=10$ and $\beta=0\textrm{.}1$, for different values of the
escape `strength' $\nu$, which is the ``transition rate'' at the escape window.
We have included for comparison the `perfect escape case', i.e. once in the
escape window the escape is instantaneous, with no dwelling time. Notice how
$\nu$ regulates the existence of a minimum in the MET: as $\nu$ gets smaller
the imperfection rises the T curve until it becomes monotonous. Hence, for this
situation we could say that $\nu$ affects negatively the `mixed' type of
transport (journey's along the boundaries and the surface).
However it is worth remarking that the transition rate at the escape window
can contribute positively as well. This behavior is well depicted in figure~\ref{Fig4}.

Figure~\ref{Fig4} presents curves corresponding to the Mean Escape Time,
as a function of the desorption rate $\delta$, with $N=10$, $M=2$, $\beta=0\textrm{.}01$,
for different values of the transition rate $\nu$.
As can be inferred from the figure, $\nu$ significantly influences MET as
it varies from $0$ to $\infty$. Changes in the location of the extrema values of MET can be seen ranging from a monotonous behavior ($\nu\rightarrow\infty$ extrema in $\delta\rightarrow\infty$) to a situation with a global minimum, and then back again into a monotonous behavior ($\nu\rightarrow0$ extrema in
$\delta\rightarrow0$). So in this case the transition rate to the escape window
contributes `positively' to the mixed type of transport, since it
turns  a situation without a minimum (perfect escape case) into a situation of
enhanced transport (minimum in T for some values of $\nu$).

\begin{figure}[!htb]
\begin{center}
\includegraphics[clip,width=0.48\textwidth]{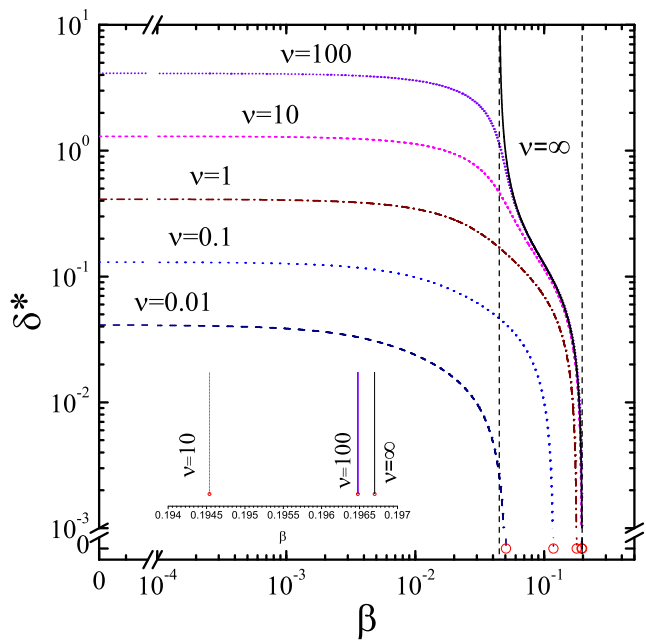}\vspace{0.0cm}
\caption{(color online) $\delta^{*}$ as a function of $\beta$ (in lin-log;lin-log scale)
for fixed $N$, $M$ and $\gamma$ for different values of $\nu$. The cuts of $\delta^*$ in
the $\beta$ axis ($\delta^{*}(\beta_{\circ})=0$) for different $\nu$, indicated by red empty
circles are not included in the curve; $\beta_{\circ}$ points are obtained from equation
(\ref{beta_lim_nu}). Vertical dashed lines are the asymptotes for the perfect escape
case obtained from equations (\ref{Asym_1}) and (\ref{Asym_2}).}
\label{Fig5}
\end{center}
\end{figure}
In Fig.\ref{Fig5} we present curves corresponding to the $\delta$ value
that minimizes MET , $\delta^*$,  as a function of $\beta$ for different
values  of $\nu$, obtained from the numerical solution of Eq. \eqref{delta_min}.
All lines depicts quite a similar trend for finite $\nu$;
$\delta^*(\beta_{\circ},N,M,\gamma,\nu)=0$ values marked by empty circles
are not included in the curves and mark the end of the $\beta$-interval in
which $\delta^*$ exists. In other words $T$ is not monotonous while
$\beta\in[0,\beta_{\circ})$. As we show in Appendix
(\ref{AppendixB}) $\beta_{\circ}$ satisfies,
\begin{equation}\label{beta_lim_nu}
2\beta_{\circ}=\frac{\gamma}{M}\frac{\sum_{q=1}^{N-1}\alpha_1\alpha_2^{-2}}{\frac{(N-1)2\beta_{\circ}}{\nu}+\sum_{q=1}^{N-1}\alpha_2^{-1}}
\end{equation}
For values larger than $\beta_{\circ}$ the T curve reaches a minimum at
$\delta=0$. However this is found at the beginning of the $\delta$-interval and without change of sign of $\partial T/\partial\delta$. We decided to rule it
out as long as it doesn't represent a true interplay between surface and boundary paths. In these situations all particles stay in the base line
and eventually escape trough the escape window without excursions into the surface.

The behavior of $\delta^*$ considerably changes in the perfect case ($\nu=\infty$).
Particularly the range of $\beta$ values where a minimum exists in T shrinks as
indicated by the dashed asymptotes in
the figure. The left/right asymptotes indicate the limit in which T
becomes monotonous, extrema for $\delta^*\rightarrow\infty$ and
$\delta^*\rightarrow0$ respectively. The left and right asymptotes are
respectively located at,
\begin{eqnarray}
\label{Asym_1}2\beta_{\delta^*\rightarrow\infty}&=&\frac{\gamma}{M}\sum_{q=1}^{N-1}\alpha_1^{-1}\left(\sum_{q=1}^{N-1}\alpha_2\alpha_1^{-2}\right)^{-1}\\
\label{Asym_2}2\beta_{\delta^*\rightarrow0}&=&\frac{\gamma}{M}\sum_{q=1}^{N-1}\alpha_1\alpha_2^{-2}\left(\sum_{q=1}^{N-1}\alpha_2^{-1}\right)^{-1}
\end{eqnarray}
For clarity's sake in the inset we have magnified the entry points to the
$\beta$ axis of $\delta^*$ curves for $\nu=10, 100, \infty$.

\begin{figure}[!htb]
\begin{center}
\includegraphics[clip,width=0.48\textwidth]{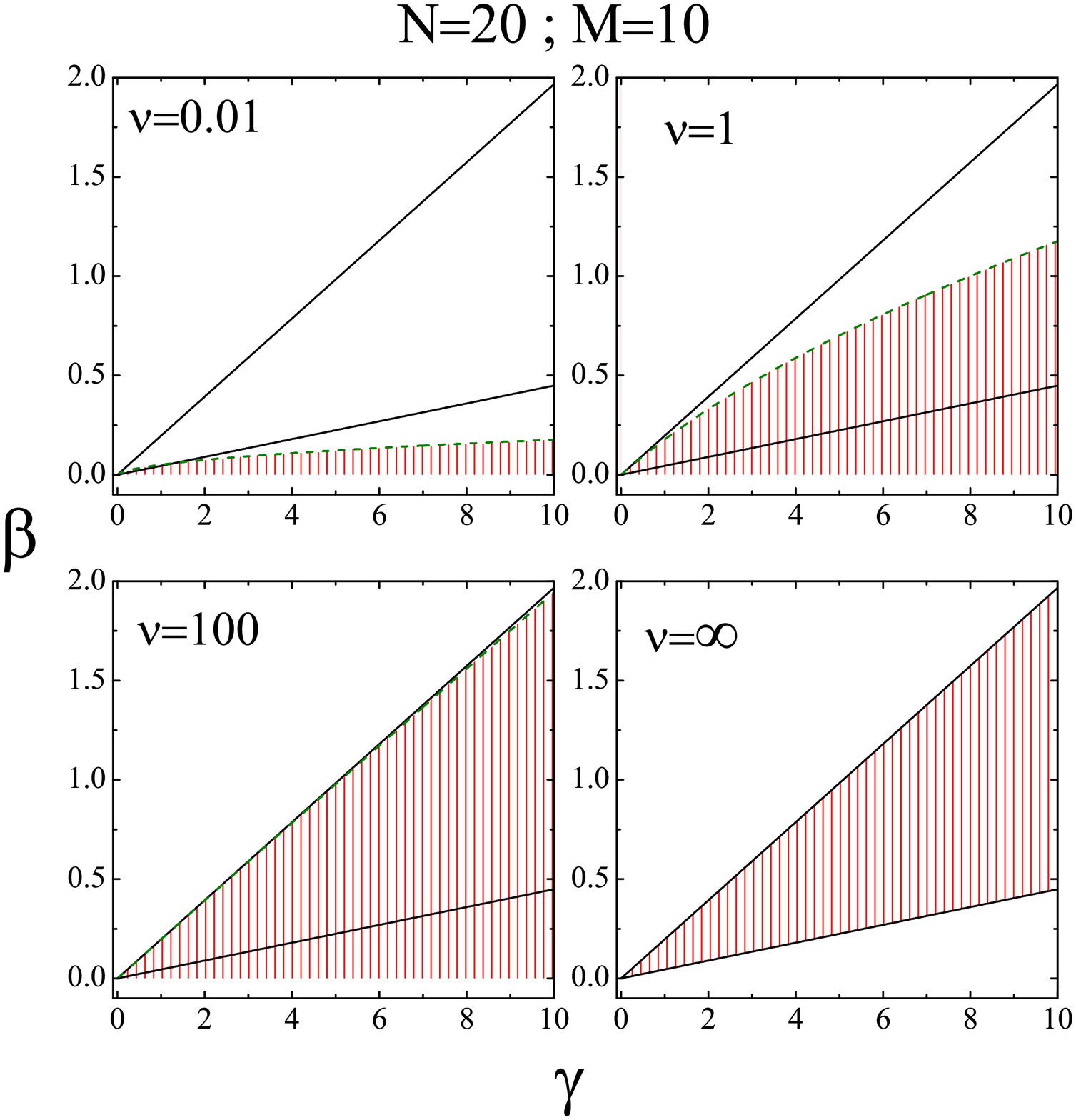}\vspace{0.0cm}
\caption{(color online) Phase diagrams that summarize the \emph{existence/non-existence}
of enhanced transport for $\nu=0\textrm{.}01, 1, 100, \infty$ respectively,
for fixed system sizes $N=20$ and $M=10$. White regions correspond to \emph{non-optimal}
transport, while filled regions (red patterns) identify regimes of enhanced transport.
Region enclosed by black lines correspond to enhanced transport in the perfect trapping
case ($\nu=\infty$) while dashed (green) lines correspond to the bound -from
eq.(\ref{beta_lim_nu})- after which, T becomes monotonous.}
\label{Fig6}
\end{center}
\end{figure}
Figure \mbox{\ref{Fig6}} shows the phase diagrams that summarize the \emph{existence/non-existence} of enhanced
transport, analyzed from the perspective of the existence of a minimum in
the Mean Escape Time. The diagrams are plotted for fixed $N$, $M$ and
$\nu$ as a function of the transition probability over the
baseline, $\beta$, and the surface transition probability $\gamma$. White regions correspond to \emph{non-optimal} transport (absence of minimum
-monotonous behaviour- in the MET), while filled regions (red patterns)
identify regimes with enhanced transport. We have also included in figure
\ref{Fig6}, enclosed by black lines the region corresponding to enhanced
transport in the perfect window escape case. Dashed curves are obtained from the solution of equation \ref{beta_lim_nu} with $\beta_{\circ}>0$,
\begin{equation}\label{beta_nu_sol}
2\beta_{\circ}\!\!=\!\!-\frac{\nu\sum_q\alpha_2^{-1}}{2(N-1)} +\sqrt{\gamma\frac{\nu\sum_q\alpha_2^{-1}}{M(N-1)}+\left(\frac{\nu\sum_q\alpha_1\alpha_2^{-2}}{2(N-1)}\right)^2}\,.
\end{equation}

As it was expected, when $\nu$ grows the regions approach the
perfect case and the escape, once in the window, becomes certain
and instantaneous. Notice that we obtain quite a good agreement
between the region of optimal transport, evaluated from equation (\ref{Tsol}),
and the corresponding bounds derived from relations (\ref{beta_nu_sol})
for finite $\nu$ and (\ref{Asym_1}), (\ref{Asym_2}) for $\nu=\infty$.

\section{Conclusions}

We have presented a model based on a master equation approach to the
\emph{narrow escape time} problem. In this study we introduced
a finite transition probability, $\nu$, at the narrow escape window
which allowed us to study the imperfect escape case. Varying from
$0$ to $\infty$, $\nu$ allowed the study of both extremes of the
trapping process: that of a highly deficient capture, and situations
where escape is certain (perfect trapping case).

By resorting to Dyson' technique we have obtained analytic results for
the primary quantity studied in the NET problem, the \emph{Mean Escape Time}
(MET), and we have studied its dependence in terms of the transition
(desorption) probability over (from) the surface boundary, the confining
domain dimensions, and the finite transition probability at the escape window.
Particularly we showed that the existence of a global minimum
in the NET is controlled by the `imperfection' of the escape
process. Regarding such conclusion, a very interesting result was that the
`imperfect' escape window enabled a region where T could be minimized, something the perfect case lacked.

We have also presented bounds -equations (\ref{beta_lim_nu}),
(\ref{Asym_1}) and (\ref{Asym_2})- between which an optimal minimum
value of T could be found, improving previous bounds derived in \cite{210}.
The phase diagrams introduced in the last section deserve a special word,
for not only do they present a compact summary of the situations of enhanced
transport, whenever some exist, but they also can lead to a better understanding of the relations among the parameters that characterize the system. In addition to our analytical approach, we have implemented Monte Carlo simulations, finding excellent agreement between the theoretical results and simulations.

We consider that the presented scheme is an analytically manageable model,
which could be used to study the impact of several (domain dimension,
different rates of transition, etc.) parameters in the interplay between surface and boundary pathways, and could also serve as a forerunner for the study of more general and complex systems. This work contributes to an area of growing interest, providing a more general overview of a previous
work (\cite{210}) and showing a plausible physical insight into the
surface-mediated diffusion mechanisms in the presence of an imperfect
escape window.

The current approach to the \emph{narrow escape time} problem
can be generalized in several directions: higher dimensions,``dynamical''
behaviour of the narrow escape window, non-markovian desorption, etc. All of
these aspects will be the subject of future work.


\begin{acknowledgments}
The authors thank C.\ E.\ Budde Jr. for technical assistance. Support by
CONICET and SeCyT (Universidad Nacional de C\'ordoba), Argentina, is
acknowledged. HSW acknowledges financial support from MICINN, Spain, through
Project FIS2010-18023.
\end{acknowledgments}

\appendix

\section{MET calculation}
\label{AppendixA}

In this appendix we focus on the calculation of the probability $P(n,m, t|n_0,m_0,t=0)$,
which is the building block for the Mean Escape Time. Taking the (finite) \emph{Fourier}
transform with respect to the $x$ variable and the \emph{Laplace} transform
with respect to the time $t$ in Equation (\ref{MsEq1}), we obtain:
\begin{eqnarray}
\nonumber m=0 &&\\
\nonumber uP(k,0,u)-P(k,0,t=0) &=&\gamma P(k,1,u)\\
\nonumber-(\delta&-&A_1(k))P(k,0,u)\\
\nonumber m=1 &&\\
\nonumber uP(k,1,u)-P(k,1,t=0)&=&\delta P(k,0,u)+ \gamma P(k,2,u)\\
\nonumber-(2\gamma &-& A(k))P(k,1,u)\\
\nonumber2\leq m \leq &M&-1\\
\nonumber uP(k,m,u)-P(k,m,t=0)&=&A(k)P(k,m,u) \\
\nonumber + \gamma(P(k,m&+&1,u)+P(k,m-1,u)\\
\nonumber& &-2P(k,m,u))\\
\nonumber m=M&&\\
\nonumber uP(k,M,u)-P(k,M,t=0)&=&A(k)P(k,M,u)+ \\
\nonumber \gamma P(k,M&-&1,u)-\gamma P(k,M,u) \,.\\
\label{MsEqT1}
\end{eqnarray}
Here we have defined $ A_1(k)=2\beta(\cos{k}-1)$, $A(k)=2\gamma(\cos{k}-1)$. Using
the matrix formalism, equation (\ref{MsEqT1}) can be written as
\begin{equation}\label{Dyson1}
\left[u\mathbb{I}-\mathbb{H}\right]\mathbb{P}=\mathbb{I}\,,
\end{equation}
where $\mathbb{I}$ is the identity matrix, $\mathbb{H}$ is an $(M+1)\times (M+1)$
tri-diagonal matrix with elements:
\begin{equation}\mathbb{H}=
\begin{bmatrix}
C_1            &   \gamma & 0        & \ldots & \ldots  &0\\
 \delta        &        C & \gamma   & 0      & \ldots  & 0\\
             0 &   \gamma & C        & \gamma & 0     &\vdots\\
 \ldots        & 0        &\ddots    & \ddots &  \ddots &  \vdots\\
     \ldots   & \ldots   & \ldots & \gamma  & C      & \gamma\\
       0 & \ldots         &        &   0     & \gamma & \gamma+C\\
      \end{bmatrix}\,,
\end{equation}
$C$ and $C_1$ are defined as $C=-2\gamma+A(k)$, $C_1=-\delta+ A_1(k)$,
and $\mathbb{P}$ is an $(M+1)\times (M+1)$ matrix with components,
\begin{displaymath}
\left[\mathbb{P}(k,u)\right]_{m,m_0}=P(k,m,u|n_0,m_0,t=0)\, .
\end{displaymath}
In order to find the solution to equation (\ref{Dyson1}) we decompose
the $\mathbb{H}$ matrix in the following way,
\begin{equation}
\mathbb{H}=A(k)\mathbb{I}+\mathbb{H}^{0}+\mathbb{H}^{1}+\mathbb{H}^{2} \,,
\end{equation}
where
\begin{equation}\mathbb{H}^0=
\begin{bmatrix}
-\gamma &   \gamma &      0   & .. & 0\\
 \gamma & -2\gamma & \gamma   & .. & 0\\
      0 &   \gamma & -2\gamma & \gamma & 0 \\
      .. & .. & .. & ..\\
      ..& .. &  \gamma & -2\gamma & \gamma\\
      ..& &    0 & \gamma & -\gamma\\
      \end{bmatrix}\,,
\end{equation}
corresponds to the transition matrix for a symmetric random walk
to nearest neighbours in a finite lattice ($M + 1$ sites) with
reflective boundary conditions at the ends. On the other hand:
\begin{eqnarray}
\mathbb{H}^1&=&(\gamma-\delta+A_1(k)\!-\!A(k))\delta_{i,0}\delta_{0,j}\!=\!\Delta_1\delta_{i,0}\delta_{0,j}\\
\mathbb{H}^2&=&-(\gamma-\delta)\,\delta_{i,1}\,\delta_{0,j}=\Delta_2\,\delta_{i,1}\,\delta_{0,j}
\end{eqnarray}

A formal solution to equation (\ref{Dyson1}) is:
\begin{equation}\label{PsolGen}
\mathbb{P}=\left[u\mathbb{I}-\mathbb{H}\right]^{-1}.
\end{equation}
By applying the Dyson procedure (\cite{050}) a general expression
-in the \emph{Fourier-Laplace} space- for $\left[\mathbb{P}(k,u)
\right]_{m,m_0}$ could be found,

\begin{widetext}
\begin{eqnarray}\label{Pmm0}
\nonumber\left[\mathbb{P}(k,u)\right]_{m,m_0}&=&\left[\mathbb{P}^0(k,u)\right]_{m,m_0}+\left[\mathbb{P}^0(k,u)\right]_{m,0}\left[\mathbb{P}^0(k,u)\right]_{0,m_0}\cdot\frac{\Delta_1}{1-\Delta_1\left[\mathbb{P}^0(k,u)\right]_{0,0}}\\
&&+ \frac{\left[\mathbb{P}^0(k,u)\right]_{0,m0}\cdot\Delta_2}{1-(\Delta_1+\Delta_2)\left[\mathbb{P}^0(k,u)\right]_{0,0}}%
\left(\left[\mathbb{P}^0(k,u)\right]_{m,1}+\frac{\left[\mathbb{P}^0(k,u)\right]_{m,0}\left[\mathbb{P}^0(k,u)\right]_{0,1}\cdot\Delta_1}{1-\Delta_1\left[\mathbb{P}^0(k,u)\right]_{0,0}}\right)
\end{eqnarray}
\end{widetext}
where,
\begin{equation}
\nonumber\left[\mathbb{P}^0(k,u)\right]_{m,m_0}=\frac{\eta^{|m-m_0|}+\eta^{\tilde{M}-(m+m_0)}}
{2\gamma(1-\eta)+\tilde{u}}\, ,\\
\end{equation}
$\tilde{u}=u-A(k)$, $\tilde{M}=2M+1$ and $\eta=1+\left(\tilde{u}-\sqrt{\tilde{u}^2+4\gamma\tilde{u}}\right)/2\gamma$.

From (\ref{Pmm0}), the probability that a walker is  at site $(n, m)$
at time $t$ given it was at $(n_0, m_0)$ at $t = 0$, $P(n,m,t|n_0,m_0,t=0)$ is
derived by using the inverse \emph{Laplace} transform on $u$ and the
inverse \emph{Fourier} transform on $k$ (for the $x$ coordinate) for
each matrix element $\left[\mathbb{P}(k,u)\right]_{m,m_0}$. Notice that,
as we are interested in the calculation of (\ref{Tdef}), we only need
to perform the inverse Fourier transform on $P(0,0,u|n_0,m_0,t=0)$ i.e.,
we need the elements $\mathcal{F}^{-1}\left\{\left[\mathbb{P}(k,u)
\right]_{0,m_0}\right\}$. In this case expression (\ref{Pmm0}) reduces to,
\begin{eqnarray}
\nonumber\left[\mathbb{P}(k,u)\right]_{0,m_0}\!\!=\!\!\frac{\eta^{m_0}+\eta^{\tilde{M}-m_0}}
{\delta(1-\eta)(1-\eta^{\tilde{M}-1})+(u-A_1(k))(1+\eta^{\tilde{M}})}\\
\end{eqnarray}
The inverse \emph{Fourier} transform on $\left[\mathbb{P}(k,u) \right]_{0,m_0}$ is carried out in the following way,
\begin{equation}
P(0,0,u|n_0,m_0,t=0)=\frac{1}{N}\sum_{q=0}^{N-1}e^{i\frac{2\pi n_0
q}{N}}\big[\mathbb{P}(\frac{2\pi q}{N},u)\big]_{0,m_0}\,.
\end{equation}

Thus we have obtained  all the required expressions for the
calculation of the MET. We now proceed to evaluate the
Mean Escape Time for a walker with an uniform initial
distribution on the base line ($y=0$), i.e.
$g(n,m)=(1-\delta_{n,0})\delta_{m,0}/(N-1)$. Notice that we
explicitly exclude the possibility of having a walker at $(0,0)$ at $t = 0$. We obtain,
\begin{equation}\label{Tsol_apA}
T=N\left[\frac{M}{\gamma}+\frac{1}{\delta}\right]\left\{\frac{\delta}{\nu}+\frac{\delta}{N-1}\sum_{q=1}^{N-1}\left[\mathbb{P}(\frac{2\pi q}{N},u=0)\right]_{0,0}\right\}
\end{equation}
or
\begin{equation}\label{Tsol_apA}
T=N\left[\frac{M}{\gamma}+\frac{1}{\delta}\right]\left\{\frac{\delta}{\nu}+\frac{\delta}{N-1}\sum_{q=1}^{N-1}\frac{1}{\alpha_1\delta+\alpha_22\beta}\right\}\,,
\end{equation}
where
\begin{eqnarray}
\alpha_1=\alpha_1(q,N,M)&=&\frac{(1-\eta_{u=0})(1-\eta_{u=0}^{2M})}{(1+\eta_{u=0})^{2M+1}}\\
\alpha_2=\alpha_2(q,N)&=&1-\cos{\frac{2\pi}{N}q}
\end{eqnarray}
\section{$\delta^*$ - optimal desorption probability}
\label{AppendixB}
In this section we present some results regarding the desorption probability
rate value that minimizes T, $\delta^*$. For this recall
equation (\ref{delta_min}),
\begin{equation}\label{delta_min_B}
\left.\frac{\partial T}{\partial\delta}\right|_{\delta=\delta^*}\!\!=\frac{NM}{\gamma\nu}+ \frac{N}{N-1}\sum_{q=1}^{N}\frac{M\gamma^{-1}2\beta\alpha_2-\alpha_1}{(\alpha_1\delta^*+\alpha_{2}2\beta)^2}=0
\end{equation}
Let us focus on the relation between $\delta$ and $\beta$ as these are the
parameters of interest, since the former enables the transport
on the surface, and the later regulates the movement on the boundary line where the escape window is located.
Although we will keep track of all variables, it could be shown that
(\ref{delta_min_B}) can be written in terms of the scaled
variables $\beta'=\beta\gamma^{-1}$, $\nu'=\nu\gamma^{-1}$ and
$\delta'=\delta\gamma^{-1}$. So a modification in $\gamma$ would result in an
enlargement or shrinkage (if we let $\gamma$ get smaller or larger respectively)
of the former variables. We will consider the behaviour of $\partial\delta^*/
\partial\beta$. Even though we were not able to obtain an explicit expression for $\delta^*$, its derivate with respect to $\beta$ could be evaluated in closed form. To do this recall equation \eqref{delta_min_B} and
differentiate it with respect to $\beta$ considering $\delta^*=f(\beta)$
(with the other parameters held fixed). Then after some algebra we
obtain,
\begin{equation}\label{dif_delta_beta}
\frac{\partial\delta^*}{\partial\beta}=\frac{\sum_{q=1}^{N-1}\frac{2(M\gamma^{-1}\delta^*+1)\alpha_1\alpha_2}{(\alpha_1\delta^*+\alpha_22\beta)^3}-\sum_{q=1}^{N-1}\frac{M\gamma^{-1}\alpha_2}{(\alpha_1\delta^*+\alpha_22\beta)^2}}{\sum_{q=1}^{N-1}\frac{(M\gamma^{-1}\delta^*+1)\alpha_1^2}{(\alpha_1\delta^*+\alpha_22\beta)^3}-\sum_{q=1}^{N-1}\frac{M\gamma^{-1}\alpha_1}{(\alpha_1\delta^*+\alpha_22\beta)^2}}
\end{equation}
From (\ref{dif_delta_beta}) we could obtain the entry points
to the $\beta$ axis of $\delta^*$ curves. The forerunner for this is the sharp
growth on the $\delta^*$ curves in figure (\ref{Fig5}). As a
matter of fact the abrupt increase is in $\partial\log{\delta^*}/\partial\log{\beta}$. However
is not difficult to show that this happens only if the denominator
in (\ref{dif_delta_beta}) $\rightarrow0$ for some $\beta_{\circ}$
(also notice that in this situation $\delta^*\sim0$) so,
\begin{eqnarray}\label{beta_lim_nuB}
\sum_{q=1}^{N-1}\frac{\alpha_1^2}{(\alpha_{2}2\beta_{\circ})^3}&=&M\gamma^{-1}\sum_{q=1}^{N-1}\frac{\alpha_1}{(\alpha_{2}2\beta_{\circ})^2}\\
\nonumber&=&(M\gamma^{-1})^2\left(\frac{(N-1)}{\nu}+2\beta_{\circ}\sum_{q=1}^{N-1}\frac{\alpha_1}{(\alpha_{2}2\beta_{\circ})^2}\right)
\end{eqnarray}
where we have used (\ref{delta_min_B}) to replace the sum on the right hand side.
We could go even further and replace the sum on the left hand side. For this we differentiate (\ref{delta_min_B}) with respect to $\delta^*$
and get  $M\gamma^{-1}2\beta\sum_q\alpha_1\alpha_2(\alpha_1\delta^*+\alpha_22\beta)^{-3}=
\sum_q\alpha_1^2(\alpha_1\delta^*+\alpha_22\beta)^{-3}$. By using this relation
and rearranging some terms in (\ref{beta_lim_nuB}) we finally obtain,
\begin{equation}\label{beta_lim_nuB_fin}
2\beta_{\circ}=\frac{\gamma}{M}\frac{\sum_{q=1}^{N-1}\alpha_1\alpha_2^{-2}}{\frac{(N-1)2\beta_{\circ}}{\nu}+\sum_{q=1}^{N-1}\alpha_2^{-1}}
\end{equation}
The solution of equation (\ref{beta_lim_nuB_fin}) in terms of $\beta_{\circ}$
that makes $\delta^*(\beta_{\circ},N,M,\gamma,\nu)=0$ marks
the end of the interval in which $\delta^*$ exists i.e., the T curve becomes
monotonous.

For the perfect escape case we obtain one of the asymptotes between
which $\delta^*$ exits, outside them the extrema is pushed either to
$0$ or to $\infty$, by letting $\nu\rightarrow\infty$ in (\ref{beta_lim_nuB_fin}),
\begin{equation}\label{beta_lim_inf_B1}
2\beta_{\delta^*\rightarrow0}=\frac{\gamma}{M}\frac{\sum_{q=1}^{N-1}\alpha_1\alpha_2^{-2}}{\sum_{q=1}^{N-1}\alpha_2^{-1}}\,,
\end{equation}
For the second asymptote we go back to (\ref{dif_delta_beta}), follow a similar
reasoning that lead us to equation (\ref{beta_lim_nuB}), here $\nu=\infty$,
$\delta^*\rightarrow\infty$, and obtain,
\begin{equation}\label{beta_lim_inf_B2}
2\beta_{\delta^*\rightarrow\infty}=\frac{\gamma}{M}\frac{\sum_{q=1}^{N-1}\alpha_1^{-1}}{\sum_{q=1}^{N-1}\alpha_2\alpha_1^{-2}}\,.
\end{equation}
Equations (\ref{beta_lim_inf_B1}) and (\ref{beta_lim_inf_B2}) constitute
an improvement to the bounds derived in \cite{210} and the solution
of equation (\ref{beta_lim_nuB_fin}) gives a bound (we could not find
similar results in the literature known by us) regarding the existence
of a minimum in the MET.

%

\begin{thebibliography}{34}%
\makeatletter
\providecommand \@ifxundefined [1]{%
 \@ifx{#1\undefined}
}%
\providecommand \@ifnum [1]{%
 \ifnum #1\expandafter \@firstoftwo
 \else \expandafter \@secondoftwo
 \fi
}%
\providecommand \@ifx [1]{%
 \ifx #1\expandafter \@firstoftwo
 \else \expandafter \@secondoftwo
 \fi
}%
\providecommand \natexlab [1]{#1}%
\providecommand \enquote  [1]{``#1''}%
\providecommand \bibnamefont  [1]{#1}%
\providecommand \bibfnamefont [1]{#1}%
\providecommand \citenamefont [1]{#1}%
\providecommand \href@noop [0]{\@secondoftwo}%
\providecommand \href [0]{\begingroup \@sanitize@url \@href}%
\providecommand \@href[1]{\@@startlink{#1}\@@href}%
\providecommand \@@href[1]{\endgroup#1\@@endlink}%
\providecommand \@sanitize@url [0]{\catcode `\\12\catcode `\$12\catcode
  `\&12\catcode `\#12\catcode `\^12\catcode `\_12\catcode `\%12\relax}%
\providecommand \@@startlink[1]{}%
\providecommand \@@endlink[0]{}%
\providecommand \url  [0]{\begingroup\@sanitize@url \@url }%
\providecommand \@url [1]{\endgroup\@href {#1}{\urlprefix }}%
\providecommand \urlprefix  [0]{URL }%
\providecommand \Eprint [0]{\href }%
\providecommand \doibase [0]{http://dx.doi.org/}%
\providecommand \selectlanguage [0]{\@gobble}%
\providecommand \bibinfo  [0]{\@secondoftwo}%
\providecommand \bibfield  [0]{\@secondoftwo}%
\providecommand \translation [1]{[#1]}%
\providecommand \BibitemOpen [0]{}%
\providecommand \bibitemStop [0]{}%
\providecommand \bibitemNoStop [0]{.\EOS\space}%
\providecommand \EOS [0]{\spacefactor3000\relax}%
\providecommand \BibitemShut  [1]{\csname bibitem#1\endcsname}%
\let\auto@bib@innerbib\@empty
\bibitem [{\citenamefont {Tka{\v{c}}ik}\ and\ \citenamefont
  {Bialek}(2009)}]{020}%
  \BibitemOpen
  \bibfield  {author} {\bibinfo {author} {\bibfnamefont {G.}~\bibnamefont
  {Tka{\v{c}}ik}}\ and\ \bibinfo {author} {\bibfnamefont {W.}~\bibnamefont
  {Bialek}},\ }\href {\doibase 10.1103/PhysRevE.79.051901} {\bibfield
  {journal} {\bibinfo  {journal} {Phys. Rev. E}\ }\textbf {\bibinfo {volume}
  {79}},\ \bibinfo {pages} {051901} (\bibinfo {year} {2009})}\BibitemShut
  {NoStop}%
\bibitem [{\citenamefont {Lomholt}\ \emph {et~al.}(2009)\citenamefont
  {Lomholt}, \citenamefont {van~den Broek}, \citenamefont {Kalisch},
  \citenamefont {Wuite},\ and\ \citenamefont {Metzler}}]{030}%
  \BibitemOpen
  \bibfield  {author} {\bibinfo {author} {\bibfnamefont {M.~A.}\ \bibnamefont
  {Lomholt}}, \bibinfo {author} {\bibfnamefont {B.}~\bibnamefont {van~den
  Broek}}, \bibinfo {author} {\bibfnamefont {S.-M.~J.}\ \bibnamefont
  {Kalisch}}, \bibinfo {author} {\bibfnamefont {G.~J.~L.}\ \bibnamefont
  {Wuite}}, \ and\ \bibinfo {author} {\bibfnamefont {R.}~\bibnamefont
  {Metzler}},\ }\href {\doibase 10.1073/pnas.0903293106} {\bibfield  {journal}
  {\bibinfo  {journal} {Proc. Natl. Acad. Sci.}\ }\textbf {\bibinfo {volume}
  {106}},\ \bibinfo {pages} {8204} (\bibinfo {year} {2009})}\BibitemShut
  {NoStop}%
\bibitem [{\citenamefont {B\'enichou}\ \emph {et~al.}(2008)\citenamefont
  {B\'enichou}, \citenamefont {Loverdo},\ and\ \citenamefont
  {Voituriez}}]{040}%
  \BibitemOpen
  \bibfield  {author} {\bibinfo {author} {\bibfnamefont {O.}~\bibnamefont
  {B\'enichou}}, \bibinfo {author} {\bibfnamefont {C.}~\bibnamefont {Loverdo}},
  \ and\ \bibinfo {author} {\bibfnamefont {R.}~\bibnamefont {Voituriez}},\
  }\href {http://stacks.iop.org/0295-5075/84/i=3/a=38003} {\bibfield  {journal}
  {\bibinfo  {journal} {EPL}\ }\textbf {\bibinfo {volume} {84}},\ \bibinfo
  {pages} {38003} (\bibinfo {year} {2008})}\BibitemShut {NoStop}%
\bibitem [{\citenamefont {da~Luz}\ \emph {et~al.}(2009)\citenamefont {da~Luz},
  \citenamefont {Grosberg}, \citenamefont {Raposo},\ and\ \citenamefont
  {Viswanathan}}]{000}%
  \BibitemOpen
  \bibfield  {author} {\bibinfo {author} {\bibfnamefont {M.~G.~E.}\
  \bibnamefont {da~Luz}}, \bibinfo {author} {\bibfnamefont {A.}~\bibnamefont
  {Grosberg}}, \bibinfo {author} {\bibfnamefont {E.~P.}\ \bibnamefont
  {Raposo}}, \ and\ \bibinfo {author} {\bibfnamefont {G.~M.}\ \bibnamefont
  {Viswanathan}},\ }\href {http://stacks.iop.org/1751-8121/42/i=43/a=430301}
  {\bibfield  {journal} {\bibinfo  {journal} {J. Phys. A: Math. Theor.}\
  }\textbf {\bibinfo {volume} {42}},\ \bibinfo {pages} {430301} (\bibinfo
  {year} {2009})}\BibitemShut {NoStop}%
\bibitem [{\citenamefont {de~Gennes}(1987)}]{055}%
  \BibitemOpen
  \bibfield  {author} {\bibinfo {author} {\bibfnamefont {P.~G.}\ \bibnamefont
  {de~Gennes}},\ }\href {\doibase DOI: 10.1016/0001-8686(87)85003-0} {\bibfield
   {journal} {\bibinfo  {journal} {Advances in Colloid and Interface Science}\
  }\textbf {\bibinfo {volume} {27}},\ \bibinfo {pages} {189 } (\bibinfo {year}
  {1987})}\BibitemShut {NoStop}%
\bibitem [{\citenamefont {Douglas}\ \emph {et~al.}(1993)\citenamefont
  {Douglas}, \citenamefont {Johnson},\ and\ \citenamefont {Granick}}]{060}%
  \BibitemOpen
  \bibfield  {author} {\bibinfo {author} {\bibfnamefont {J.~F.}\ \bibnamefont
  {Douglas}}, \bibinfo {author} {\bibfnamefont {H.~E.}\ \bibnamefont
  {Johnson}}, \ and\ \bibinfo {author} {\bibfnamefont {S.}~\bibnamefont
  {Granick}},\ }\href {\doibase 10.1126/science.262.5142.2010} {\bibfield
  {journal} {\bibinfo  {journal} {Science}\ }\textbf {\bibinfo {volume}
  {262}},\ \bibinfo {pages} {2010} (\bibinfo {year} {1993})}\BibitemShut
  {NoStop}%
\bibitem [{\citenamefont {Bychuk}\ and\ \citenamefont
  {O'Shaughnessy}(1995)}]{065}%
  \BibitemOpen
  \bibfield  {author} {\bibinfo {author} {\bibfnamefont {O.~V.}\ \bibnamefont
  {Bychuk}}\ and\ \bibinfo {author} {\bibfnamefont {B.}~\bibnamefont
  {O'Shaughnessy}},\ }\href {\doibase 10.1103/PhysRevLett.74.1795} {\bibfield
  {journal} {\bibinfo  {journal} {Phys. Rev. Lett.}\ }\textbf {\bibinfo
  {volume} {74}},\ \bibinfo {pages} {1795} (\bibinfo {year}
  {1995})}\BibitemShut {NoStop}%
\bibitem [{\citenamefont {Clint}(1992)}]{066}%
  \BibitemOpen
  \bibfield  {author} {\bibinfo {author} {\bibfnamefont {J.~H.}\ \bibnamefont
  {Clint}},\ }\href@noop {} {\emph {\bibinfo {title} {{Surfactant
  Aggregation}}}}\ (\bibinfo  {publisher} {Chapman and Hall, New York},\
  \bibinfo {year} {1992})\BibitemShut {NoStop}%
\bibitem [{\citenamefont {Berg}\ and\ \citenamefont {Purcell}(1977)}]{070}%
  \BibitemOpen
  \bibfield  {author} {\bibinfo {author} {\bibfnamefont {H.}~\bibnamefont
  {Berg}}\ and\ \bibinfo {author} {\bibfnamefont {E.}~\bibnamefont {Purcell}},\
  }\href {\doibase 10.1016/S0006-3495(77)85544-6} {\bibfield  {journal}
  {\bibinfo  {journal} {Biophysical}\ }\textbf {\bibinfo {volume} {20}},\
  \bibinfo {pages} {193} (\bibinfo {year} {1977})}\BibitemShut {NoStop}%
\bibitem [{\citenamefont {Berg}\ \emph {et~al.}(1981)\citenamefont {Berg},
  \citenamefont {Winter},\ and\ \citenamefont {Von~Hippel}}]{071}%
  \BibitemOpen
  \bibfield  {author} {\bibinfo {author} {\bibfnamefont {O.~G.}\ \bibnamefont
  {Berg}}, \bibinfo {author} {\bibfnamefont {R.~B.}\ \bibnamefont {Winter}}, \
  and\ \bibinfo {author} {\bibfnamefont {P.~H.}\ \bibnamefont {Von~Hippel}},\
  }\href {\doibase 10.1021/bi00527a028} {\bibfield  {journal} {\bibinfo
  {journal} {Biochemistry}\ }\textbf {\bibinfo {volume} {20}},\ \bibinfo
  {pages} {6929} (\bibinfo {year} {1981})}\BibitemShut {NoStop}%
\bibitem [{\citenamefont {Linderman}\ and\ \citenamefont
  {Lauffenburger}(1986)}]{073}%
  \BibitemOpen
  \bibfield  {author} {\bibinfo {author} {\bibfnamefont {J.}~\bibnamefont
  {Linderman}}\ and\ \bibinfo {author} {\bibfnamefont {D.}~\bibnamefont
  {Lauffenburger}},\ }\href {\doibase 10.1016/S0006-3495(86)83463-4} {\bibfield
   {journal} {\bibinfo  {journal} {Biophysical}\ }\textbf {\bibinfo {volume}
  {50}},\ \bibinfo {pages} {295} (\bibinfo {year} {1986})}\BibitemShut
  {NoStop}%
\bibitem [{\citenamefont {Zhou}\ and\ \citenamefont {Zwanzig}(1991)}]{075}%
  \BibitemOpen
  \bibfield  {author} {\bibinfo {author} {\bibfnamefont {H.-X.}\ \bibnamefont
  {Zhou}}\ and\ \bibinfo {author} {\bibfnamefont {R.}~\bibnamefont {Zwanzig}},\
  }\href {\doibase 10.1063/1.460427} {\bibfield  {journal} {\bibinfo  {journal}
  {The Journal of Chemical Physics}\ }\textbf {\bibinfo {volume} {94}},\
  \bibinfo {pages} {6147} (\bibinfo {year} {1991})}\BibitemShut {NoStop}%
\bibitem [{\citenamefont {Grigoriev}\ \emph {et~al.}(2002)\citenamefont
  {Grigoriev}, \citenamefont {Makhnovskii}, \citenamefont {Berezhkovskii},\
  and\ \citenamefont {Zitserman}}]{078}%
  \BibitemOpen
  \bibfield  {author} {\bibinfo {author} {\bibfnamefont {I.~V.}\ \bibnamefont
  {Grigoriev}}, \bibinfo {author} {\bibfnamefont {Y.~A.}\ \bibnamefont
  {Makhnovskii}}, \bibinfo {author} {\bibfnamefont {A.~M.}\ \bibnamefont
  {Berezhkovskii}}, \ and\ \bibinfo {author} {\bibfnamefont {V.~Y.}\
  \bibnamefont {Zitserman}},\ }\href {http://link.aip.org/link/?JCP/116/9574/1}
  {\bibfield  {journal} {\bibinfo  {journal} {The Journal of Chemical Physics}\
  }\textbf {\bibinfo {volume} {116}},\ \bibinfo {pages} {9574} (\bibinfo {year}
  {2002})}\BibitemShut {NoStop}%
\bibitem [{\citenamefont {Singer}\ \emph
  {et~al.}(2006{\natexlab{a}})\citenamefont {Singer}, \citenamefont {Schuss},
  \citenamefont {Holcman},\ and\ \citenamefont {Eisenberg}}]{079}%
  \BibitemOpen
  \bibfield  {author} {\bibinfo {author} {\bibfnamefont {A.}~\bibnamefont
  {Singer}}, \bibinfo {author} {\bibfnamefont {Z.}~\bibnamefont {Schuss}},
  \bibinfo {author} {\bibfnamefont {D.}~\bibnamefont {Holcman}}, \ and\
  \bibinfo {author} {\bibfnamefont {R.}~\bibnamefont {Eisenberg}},\ }\href@noop
  {} {\bibfield  {journal} {\bibinfo  {journal} {Journal of Statistical
  Physics}\ }\textbf {\bibinfo {volume} {122}},\ \bibinfo {pages} {437}
  (\bibinfo {year} {2006}{\natexlab{a}})}\BibitemShut {NoStop}%
\bibitem [{\citenamefont {Singer}\ \emph
  {et~al.}(2006{\natexlab{b}})\citenamefont {Singer}, \citenamefont {Schuss},\
  and\ \citenamefont {Holcman}}]{080}%
  \BibitemOpen
  \bibfield  {author} {\bibinfo {author} {\bibfnamefont {A.}~\bibnamefont
  {Singer}}, \bibinfo {author} {\bibfnamefont {Z.}~\bibnamefont {Schuss}}, \
  and\ \bibinfo {author} {\bibfnamefont {D.}~\bibnamefont {Holcman}},\ }\href
  {http://dx.doi.org/10.1007/s10955-005-8027-5} {\bibfield  {journal} {\bibinfo
   {journal} {Journal of Statistical Physics}\ }\textbf {\bibinfo {volume}
  {122}},\ \bibinfo {pages} {465} (\bibinfo {year}
  {2006}{\natexlab{b}})}\BibitemShut {NoStop}%
\bibitem [{\citenamefont {Singer}\ and\ \citenamefont {Schuss}(2006)}]{090}%
  \BibitemOpen
  \bibfield  {author} {\bibinfo {author} {\bibfnamefont {A.}~\bibnamefont
  {Singer}}\ and\ \bibinfo {author} {\bibfnamefont {Z.}~\bibnamefont
  {Schuss}},\ }\href {\doibase 10.1103/PhysRevE.74.020103} {\bibfield
  {journal} {\bibinfo  {journal} {Phys. Rev. E}\ }\textbf {\bibinfo {volume}
  {74}},\ \bibinfo {pages} {020103} (\bibinfo {year} {2006})}\BibitemShut
  {NoStop}%
\bibitem [{\citenamefont {Bressloff}\ and\ \citenamefont
  {Earnshaw}(2007)}]{095}%
  \BibitemOpen
  \bibfield  {author} {\bibinfo {author} {\bibfnamefont {P.~C.}\ \bibnamefont
  {Bressloff}}\ and\ \bibinfo {author} {\bibfnamefont {B.~A.}\ \bibnamefont
  {Earnshaw}},\ }\href {\doibase 10.1103/PhysRevE.75.041915} {\bibfield
  {journal} {\bibinfo  {journal} {Phys. Rev. E}\ }\textbf {\bibinfo {volume}
  {75}},\ \bibinfo {pages} {041915} (\bibinfo {year} {2007})}\BibitemShut
  {NoStop}%
\bibitem [{\citenamefont {B\'enichou}\ and\ \citenamefont
  {Voituriez}(2008)}]{100}%
  \BibitemOpen
  \bibfield  {author} {\bibinfo {author} {\bibfnamefont {O.}~\bibnamefont
  {B\'enichou}}\ and\ \bibinfo {author} {\bibfnamefont {R.}~\bibnamefont
  {Voituriez}},\ }\href {\doibase 10.1103/PhysRevLett.100.168105} {\bibfield
  {journal} {\bibinfo  {journal} {Phys. Rev. Lett.}\ }\textbf {\bibinfo
  {volume} {100}},\ \bibinfo {pages} {168105} (\bibinfo {year}
  {2008})}\BibitemShut {NoStop}%
\bibitem [{\citenamefont {Oshanin}\ \emph {et~al.}(2010)\citenamefont
  {Oshanin}, \citenamefont {Tamm},\ and\ \citenamefont {Vasilyev}}]{110}%
  \BibitemOpen
  \bibfield  {author} {\bibinfo {author} {\bibfnamefont {G.}~\bibnamefont
  {Oshanin}}, \bibinfo {author} {\bibfnamefont {M.}~\bibnamefont {Tamm}}, \
  and\ \bibinfo {author} {\bibfnamefont {O.}~\bibnamefont {Vasilyev}},\ }\href
  {\doibase 10.1063/1.3442906} {\bibfield  {journal} {\bibinfo  {journal} {The
  Journal of Chemical Physics}\ }\textbf {\bibinfo {volume} {132}},\ \bibinfo
  {eid} {235101} (\bibinfo {year} {2010})}\BibitemShut {NoStop}%
\bibitem [{\citenamefont {B\'enichou}\ \emph {et~al.}(2010)\citenamefont
  {B\'enichou}, \citenamefont {Grebenkov}, \citenamefont {Levitz},
  \citenamefont {Loverdo},\ and\ \citenamefont {Voituriez}}]{114}%
  \BibitemOpen
  \bibfield  {author} {\bibinfo {author} {\bibfnamefont {O.}~\bibnamefont
  {B\'enichou}}, \bibinfo {author} {\bibfnamefont {D.}~\bibnamefont
  {Grebenkov}}, \bibinfo {author} {\bibfnamefont {P.}~\bibnamefont {Levitz}},
  \bibinfo {author} {\bibfnamefont {C.}~\bibnamefont {Loverdo}}, \ and\
  \bibinfo {author} {\bibfnamefont {R.}~\bibnamefont {Voituriez}},\ }\href
  {\doibase 10.1103/PhysRevLett.105.150606} {\bibfield  {journal} {\bibinfo
  {journal} {Phys. Rev. Lett.}\ }\textbf {\bibinfo {volume} {105}},\ \bibinfo
  {pages} {150606} (\bibinfo {year} {2010})}\BibitemShut {NoStop}%
\bibitem [{\citenamefont {B\'enichou}\ \emph {et~al.}(2011)\citenamefont
  {B\'enichou}, \citenamefont {Grebenkov}, \citenamefont {Levitz},
  \citenamefont {Loverdo},\ and\ \citenamefont {Voituriez}}]{115}%
  \BibitemOpen
  \bibfield  {author} {\bibinfo {author} {\bibfnamefont {O.}~\bibnamefont
  {B\'enichou}}, \bibinfo {author} {\bibfnamefont {D.}~\bibnamefont
  {Grebenkov}}, \bibinfo {author} {\bibfnamefont {P.}~\bibnamefont {Levitz}},
  \bibinfo {author} {\bibfnamefont {C.}~\bibnamefont {Loverdo}}, \ and\
  \bibinfo {author} {\bibfnamefont {R.}~\bibnamefont {Voituriez}},\ }\href
  {http://dx.doi.org/10.1007/s10955-011-0138-6} {\bibfield  {journal} {\bibinfo
   {journal} {Journal of Statistical Physics}\ }\textbf {\bibinfo {volume}
  {142}},\ \bibinfo {pages} {657} (\bibinfo {year} {2011})}\BibitemShut
  {NoStop}%
\bibitem [{\citenamefont {Rojo}\ and\ \citenamefont {Budde}(2011)}]{210}%
  \BibitemOpen
  \bibfield  {author} {\bibinfo {author} {\bibfnamefont {F.}~\bibnamefont
  {Rojo}}\ and\ \bibinfo {author} {\bibfnamefont {C.~E.}\ \bibnamefont
  {Budde}},\ }\href {\doibase 10.1103/PhysRevE.84.021117} {\bibfield  {journal}
  {\bibinfo  {journal} {Phys. Rev. E}\ }\textbf {\bibinfo {volume} {84}},\
  \bibinfo {pages} {021117} (\bibinfo {year} {2011})}\BibitemShut {NoStop}%
\bibitem [{\citenamefont {Walder}\ \emph {et~al.}(2011)\citenamefont {Walder},
  \citenamefont {Nelson},\ and\ \citenamefont {Schwartz}}]{211}%
  \BibitemOpen
  \bibfield  {author} {\bibinfo {author} {\bibfnamefont {R.}~\bibnamefont
  {Walder}}, \bibinfo {author} {\bibfnamefont {N.}~\bibnamefont {Nelson}}, \
  and\ \bibinfo {author} {\bibfnamefont {D.~K.}\ \bibnamefont {Schwartz}},\
  }\href {\doibase 10.1103/PhysRevLett.107.156102} {\bibfield  {journal}
  {\bibinfo  {journal} {Phys. Rev. Lett.}\ }\textbf {\bibinfo {volume} {107}},\
  \bibinfo {pages} {156102} (\bibinfo {year} {2011})}\BibitemShut {NoStop}%
\bibitem [{\citenamefont {Calandre}\ \emph {et~al.}(2012)\citenamefont
  {Calandre}, \citenamefont {B\'enichou}, \citenamefont {Grebenkov},\ and\
  \citenamefont {Voituriez}}]{212}%
  \BibitemOpen
  \bibfield  {author} {\bibinfo {author} {\bibfnamefont {T.}~\bibnamefont
  {Calandre}}, \bibinfo {author} {\bibfnamefont {O.}~\bibnamefont
  {B\'enichou}}, \bibinfo {author} {\bibfnamefont {D.~S.}\ \bibnamefont
  {Grebenkov}}, \ and\ \bibinfo {author} {\bibfnamefont {R.}~\bibnamefont
  {Voituriez}},\ }\href {\doibase 10.1103/PhysRevE.85.051111} {\bibfield
  {journal} {\bibinfo  {journal} {Phys. Rev. E}\ }\textbf {\bibinfo {volume}
  {85}},\ \bibinfo {pages} {051111} (\bibinfo {year} {2012})}\BibitemShut
  {NoStop}%
\bibitem [{\citenamefont {Haus}\ and\ \citenamefont {Kehr}(1987)}]{180}%
  \BibitemOpen
  \bibfield  {author} {\bibinfo {author} {\bibfnamefont {J.~W.}\ \bibnamefont
  {Haus}}\ and\ \bibinfo {author} {\bibfnamefont {K.~W.}\ \bibnamefont
  {Kehr}},\ }\href {\doibase DOI: 10.1016/0370-1573(87)90005-6} {\bibfield
  {journal} {\bibinfo  {journal} {Physics Reports}\ }\textbf {\bibinfo {volume}
  {150}},\ \bibinfo {pages} {263 } (\bibinfo {year} {1987})}\BibitemShut
  {NoStop}%
\bibitem [{\citenamefont {R\'e}\ and\ \citenamefont {Budde}(2000)}]{190}%
  \BibitemOpen
  \bibfield  {author} {\bibinfo {author} {\bibfnamefont {M.~A.}\ \bibnamefont
  {R\'e}}\ and\ \bibinfo {author} {\bibfnamefont {C.~E.}\ \bibnamefont
  {Budde}},\ }\href {\doibase 10.1103/PhysRevE.61.1110} {\bibfield  {journal}
  {\bibinfo  {journal} {Phys. Rev. E}\ }\textbf {\bibinfo {volume} {61}},\
  \bibinfo {pages} {1110} (\bibinfo {year} {2000})}\BibitemShut {NoStop}%
\bibitem [{\citenamefont {Nadler}\ and\ \citenamefont {Stein}(1996)}]{200}%
  \BibitemOpen
  \bibfield  {author} {\bibinfo {author} {\bibfnamefont {W.}~\bibnamefont
  {Nadler}}\ and\ \bibinfo {author} {\bibfnamefont {D.~L.}\ \bibnamefont
  {Stein}},\ }\href {\doibase 10.1063/1.471810} {\bibfield  {journal} {\bibinfo
   {journal} {The Journal of Chemical Physics}\ }\textbf {\bibinfo {volume}
  {104}},\ \bibinfo {pages} {1918} (\bibinfo {year} {1996})}\BibitemShut
  {NoStop}%
\bibitem [{\citenamefont {Coppey}\ \emph {et~al.}(2004)\citenamefont {Coppey},
  \citenamefont {B\'enichou}, \citenamefont {Voituriez},\ and\ \citenamefont
  {Moreau}}]{201}%
  \BibitemOpen
  \bibfield  {author} {\bibinfo {author} {\bibfnamefont {M.}~\bibnamefont
  {Coppey}}, \bibinfo {author} {\bibfnamefont {O.}~\bibnamefont {B\'enichou}},
  \bibinfo {author} {\bibfnamefont {R.}~\bibnamefont {Voituriez}}, \ and\
  \bibinfo {author} {\bibfnamefont {M.}~\bibnamefont {Moreau}},\ }\href
  {\doibase 10.1529/biophysj.104.045773} {\bibfield  {journal} {\bibinfo
  {journal} {Biophysical Journal}\ }\textbf {\bibinfo {volume} {87}},\ \bibinfo
  {pages} {1640 } (\bibinfo {year} {2004})}\BibitemShut {NoStop}%
\bibitem [{\citenamefont {Berezhkovskii}\ and\ \citenamefont
  {Barzykin}(2012)}]{202}%
  \BibitemOpen
  \bibfield  {author} {\bibinfo {author} {\bibfnamefont {A.~M.}\ \bibnamefont
  {Berezhkovskii}}\ and\ \bibinfo {author} {\bibfnamefont {A.~V.}\ \bibnamefont
  {Barzykin}},\ }\href {\doibase 10.1063/1.3682243} {\bibfield  {journal}
  {\bibinfo  {journal} {The Journal of Chemical Physics}\ }\textbf {\bibinfo
  {volume} {136}},\ \bibinfo {eid} {054115} (\bibinfo {year}
  {2012})}\BibitemShut {NoStop}%
\bibitem [{\citenamefont {Revelli}\ \emph {et~al.}(2005)\citenamefont
  {Revelli}, \citenamefont {Budde},\ and\ \citenamefont {Wio}}]{050}%
  \BibitemOpen
  \bibfield  {author} {\bibinfo {author} {\bibfnamefont {J.~A.}\ \bibnamefont
  {Revelli}}, \bibinfo {author} {\bibfnamefont {C.~E.}\ \bibnamefont {Budde}},
  \ and\ \bibinfo {author} {\bibfnamefont {H.~S.}\ \bibnamefont {Wio}},\ }\href
  {http://stacks.iop.org/0953-8984/17/i=49/a=012} {\bibfield  {journal}
  {\bibinfo  {journal} {Journal of Physics: Condensed Matter}\ }\textbf
  {\bibinfo {volume} {17}},\ \bibinfo {pages} {S4175} (\bibinfo {year}
  {2005})}\BibitemShut {NoStop}%
\bibitem [{\citenamefont {Spouge}\ \emph {et~al.}(1996)\citenamefont {Spouge},
  \citenamefont {Szabo},\ and\ \citenamefont {Weiss}}]{220}%
  \BibitemOpen
  \bibfield  {author} {\bibinfo {author} {\bibfnamefont {J.~L.}\ \bibnamefont
  {Spouge}}, \bibinfo {author} {\bibfnamefont {A.}~\bibnamefont {Szabo}}, \
  and\ \bibinfo {author} {\bibfnamefont {G.~H.}\ \bibnamefont {Weiss}},\ }\href
  {\doibase 10.1103/PhysRevE.54.2248} {\bibfield  {journal} {\bibinfo
  {journal} {Phys. Rev. E}\ }\textbf {\bibinfo {volume} {54}},\ \bibinfo
  {pages} {2248} (\bibinfo {year} {1996})}\BibitemShut {NoStop}%
\bibitem [{\citenamefont {Rojo}\ \emph {et~al.}(2010)\citenamefont {Rojo},
  \citenamefont {Pury},\ and\ \citenamefont {Budde}}]{130}%
  \BibitemOpen
  \bibfield  {author} {\bibinfo {author} {\bibfnamefont {F.}~\bibnamefont
  {Rojo}}, \bibinfo {author} {\bibfnamefont {P.~A.}\ \bibnamefont {Pury}}, \
  and\ \bibinfo {author} {\bibfnamefont {C.~E.}\ \bibnamefont {Budde}},\ }\href
  {\doibase DOI: 10.1016/j.physa.2010.04.025} {\bibfield  {journal} {\bibinfo
  {journal} {Physica A}\ }\textbf {\bibinfo {volume} {389}},\ \bibinfo {pages}
  {3399 } (\bibinfo {year} {2010})}\BibitemShut {NoStop}%
\bibitem [{Note1()}]{Note1}%
  \BibitemOpen
  \bibinfo {note} {See \cite {180} and references therein.}\BibitemShut {Stop}%
\bibitem [{Note2()}]{Note2}%
  \BibitemOpen
  \bibinfo {note} {This consideration will help us in the comparison with the
  perfect escape case and avoids the `instantaneous' escaping in the limit $\nu
  \rightarrow \infty $}\BibitemShut {NoStop}%
\end{thebibliography}
%

\end{document}